# Volume-of-fluid simulations in microfluidic T-junction devices: Influence of viscosity ratio on droplet size

Mehdi Nekouei and Siva A. Vanapalli

*Department of Chemical Engineering, Texas Tech University, Lubbock, TX, 79409, USA*

**Abstract**

We used volume-of-fluid (VOF) method to perform three-dimensional numerical simulations of droplet formation of Newtonian fluids in microfluidic T-junction devices. To evaluate the performance of the VOF method we examined the regimes of drop formation and determined droplet size as a function of system parameters. Comparison of the simulation results with four sets of experimental data from the literature showed good agreement, validating the VOF method. Motivated by the lack of adequate studies investigating the influence of viscosity ratio (λ) on the generated droplet size, we mapped the dependence of drop volume on capillary number (0.001 < Ca < 0.5) and viscosity ratio (0.01 < λ < 15). We find that for all viscosity ratios investigated, droplet size decreases with increase in capillary number. However, the reduction in droplet size with capillary number is stronger for λ < 1 than for λ > 1. In addition, we find that at a given capillary number, the size of droplets does not vary appreciably when λ < 1, while it increases when λ > 1. We develop an analytical model for predicting droplet size that includes a viscosity-dependent breakup time for the dispersed phase. This improved model successfully predicts the effects of viscosity ratio observed in simulations. Results from this study are useful for the design of lab-on-chip technologies and manufacture of microfluidic emulsions, where there is a need to know how system parameters influence droplet size.

## 1. Introduction

Droplet-based microfluidics where fluid volumes down to picoliters are manipulated has witnessed a remarkable growth due to applications in biochemical analysis and material synthesis [1-5]. In these applications, it is necessary to produce droplets of controlled size. Droplet size is an extremely important parameter as it controls the efficiency of encapsulation of individual cells and biomolecules [6-8] and rates of reaction [1, 9, 10]. From a fundamental point of view, droplet size dictates the mixing dynamics [1, 9-13], flow resistance [14-17], breakup [18-20], coalescence [21-23] and collective [24, 25] behavior. Since experimental conditions including flow rates, fluid properties and channel dimensions, may vary for different applications [26] it is important to understand drop formation and develop predictive models of how system parameters influence droplet size.

A widely used microfluidic geometry for producing droplets is the T-junction device, where the continuous and dispersed phase flowing orthogonally meet at a junction producing droplets [27, 28]. Several parameters have been shown to influence droplet size in T-junction devices [29-32]. These include dimensionless parameters such as flow rate ratio ($Q = \frac{Q_D}{Q_C}$, where $Q_D$ and $Q_C$ are volumetric flow rates of dispersed and continuous phases, respectively), capillary number ($Ca = \frac{\mu_C U}{\gamma}$, where $\mu_C$ and U are the viscosity and velocity of the continuous phase, and γ is the interfacial tension), capillary number of dispersed phase ($Ca_D = \frac{\mu_D U_D}{\gamma}$, where $\mu_D$ is the viscosity of dispersed phase, $U_D$ is the inlet velocity of dispersed phase), Reynolds number ($Re = \frac{\rho_C U w_C}{\mu_C}$ where $\rho_C$ is the density of continuous phase and $w_C$ is the width of main channel), viscosity ratio($\lambda = \frac{\mu_D}{\mu_C}$) and density ratio ($\rho = \frac{\rho_D}{\rho_C}$, where $\rho_D$ is the density of dispersed phase). In addition, geometrical parameters such as width ratio ($W = \frac{w_D}{w_C}$, where $w_D$ is the width of side channel) and height ratio ($H = \frac{h}{w_C}$, where h is the height of the channel) can also influence droplet size.

Table 1. Experimental studies of droplet formation in microfluidic T-junction devices, whose results on droplet size have been compared to VOF simulations.

| Study | Ca | Q | λ | W | Focus of study |
|---|---|---|---|---|---|
| Garstecki *et al.*, 2006 [30] | Ca<0.01 | 0.01<Q<10 | 0.01<λ<0.1 | 0.25<W<1 | Droplet size dependency on Q |
| Van steijn *et al.*, 2012 [31] | Ca<0.01 | 0.1<Q<10 | 0.01<λ<0.1 | 0.33<W<3 | Droplet size dependency on T-junction geometry |
| Christopher *et al.*, 2008 [29] | 0.001<Ca<0.5 | 0.05<Q<4.5 | 0.003<λ<0.167 | 0.5<W<2.5 | Droplet size dependency on Ca |

Several experimental studies have investigated the dependence of drop size on system parameters in T-junction devices [29-36]. Among these, here we discuss those that are relevant to our work (see Table 1). Garstecki *et al.* focused on the effect of flow rate ratio on droplet size and found a linear relationship between drop length (L) and flow rate ratio [30]. Van Steijn *et al.* investigated how the dimensions of the main and side channel in the T-junction influence droplet size [31]. Similar to Garstecki *et al.*, for a given T-junction geometry, they also found linear dependence of drop volume (V) on flow rate ratio. However, they showed that at a given flow rate ratio, drop size also depends on width ratio (W) and height ratio (H). In both these studies, experiments were conducted at low capillary numbers (Ca < 0.01). Christopher *et al.* measured droplet size across a wider range of capillary numbers (0.001 < Ca < 0.5) while maintaining fixed flow rate ratios [29]. In addition to linear dependency of droplet size on flow rate ratio, they found that the droplet size decreases by increasing capillary number. The parameter range covered by these experimental studies is shown in Table 1.

Investigating experimentally, the effect of system parameters on droplet size has limitations. For example, experimentally it is difficult to produce droplets for viscosity ratios greater than unity, making it unclear how viscosity ratio influences droplet size. This feature is evident from Table 1, where studies were limited to viscosity ratio less than 1. Thus, fixing one control parameter while changing the other parameters is not always guaranteed in experiments, lending access to a narrow system parameter space. Moreover, it is difficult to experimentally measure the three-dimensional flow fields and fluid stresses during drop production in T-junction, although some progress has been made [37-40].

Table 2. Summary of numerical simulation studies on droplet formation in microfluidic T-junction devices.

| Study | Ca | Q | λ | W | Simulation method |
|---|---|---|---|---|---|
| Van der Graaf et al., 2006 [41] | 0.01<Ca<0.08 | 0.05<Q<1 | λ=3.44 | W=1 | LBM |
| Sang et al. 2009 [45] | 0.002<Ca<0.8 | 0.05<Q<1 | 0.05<λ0.3 | W=1 | VOF |
| Kashid et al., 2010[42] | 0.0047<Ca<0.012 | Q=0.5 | 0.002<λ<0.018 | W=1 | VOF, ANSYS Fluent |
| De Menech et al., 2008 [38] | 0.001<Ca<0.07 | 0.01<Q<2 | 0.125<λ<1 | W=1 | Phase field |
| Liu et al., 2009 [43] | 0.003<Ca<0.06 | 0.1<Q<1 | 0.0125<λ<1 | W=1 | LBM |
| Sivasamy et al., 2011[37] | 0.008<Ca<0.025 | Not reported | λ~0.04 | W=0.5 | VOF, ANSYS Fluent |
| Yang et al., 2013 [39] | 0.002<Ca<0.056 | 0.125<Q<1 | λ=1 | W=1 | LBM |
| Hoang et al., 2013 [44] | Ca<0.01 | 1<Q<4 | λ~0.01 | W=1 | VOF, OpenFOAM |
| Present study | 0.001<Ca<0.5 | 0.05<Q<10 | 0.01<λ<15 | 0.33<W<3 | VOF, OpenFOAM |

Numerical simulations provide a unique opportunity to complement experimental investigations. As shown in Table 2, several numerical simulations have been pursued to investigate droplet production in T-junction geometries [37-45]. These multiphase simulation approaches include volume-of-fluid (VOF) [46-48], phase field [49-52] and lattice-Boltzmann methods (LBM) [53-56]. Most of the numerical simulations to date have been focused on investigating the mechanism of drop formation at T-junction [37-40, 42, 43]. These studies showed that as the dispersed phase enters and fills the main channel, the upstream pressure increases due to blockage of the channel. This built-up pressure along with shear force exerted on dispersed phase causes the break-up of the dispersed phase. Furthermore, it was found that at higher capillary numbers, the contribution of shear force to drop formation increases, while built-up pressure decreases.

Despite several experimental studies and numerical simulations investigating droplet production in T-junction devices, additional studies are warranted due to the following reasons. First, even though simulation efforts have provided useful insights into the mechanism of drop formation in T-junction geometries, much less attention has been devoted to comprehensive investigations of the influence of system parameters on droplet size. Specifically, simulations targeting direct comparison of different sets of experimental data on droplet size has not been pursued to date. Second, the role of fluid viscosities on droplet size has not been fully elucidated. In both experiments and simulations (c.f. Table 2), so far studies have been limited to viscosity ratio less than unity. Given that for λ > 1, the viscous stress of dispersed phase can impact drop production, studies covering a broader range of viscosity ratios need to be pursued.

In this study, we use numerical simulations based on VOF method to study the influence of system parameters on drop size (see Table 2). In the first part of our study, we carried out numerical simulations to predict regimes of drop formation and the generated droplet size. We compared the simulation results with data from experimental studies shown in Table 1, allowing us to assess the capability of VOF method and validate it. In the second part of the study, we investigated the influence of viscosity ratio on droplet size and find it to play an important role. Subsequently, we develop a model that predicts droplet size considering the effect of viscosity ratio.

## 2. Numerical Simulation

### 2.1. Volume-of-fluid method and its implementation

Three-dimensional simulation of droplet formation in T-junction geometries was performed using volume-of-fluid (VOF) method. VOF is a Eulerian method of multiphase flow simulations where fluid properties such as viscosity and density are smoothed and the surface tension force is distributed over a thin layer near the interface as a body force. In VOF, a phase fraction parameter, α, is used to indicate the presence of each phase at every location of the simulation domain. In our simulation, α =1 for phase 1 (i.e. continuous phase), α = 0 for phase 2 (i.e. dispersed phase) and 0<α<1 in the interface region. In VOF, the governing equations including continuity (Eqn.1), momentum balance (Eqn. 2) and phase fraction equations (Eqn. 3), are solved simultaneously.

$$\nabla \cdot \boldsymbol{U} = 0 \quad (1)$$

$$\frac{\partial \rho_b \boldsymbol{U}}{\partial \bar{t}} + \nabla \cdot (\rho_b \boldsymbol{U}\boldsymbol{U}) = -\nabla p + \nabla \cdot \boldsymbol{T} + \rho_b \boldsymbol{f} + \boldsymbol{F}_s \quad (2)$$

$$\frac{\partial \alpha}{\partial \bar{t}} + \nabla \cdot (\alpha \boldsymbol{U}) = 0 \quad (3)$$

In Eqns. 1-3, **U** is the velocity vector field, p is the pressure field, **T** is the deviatoric stress tensor ($\boldsymbol{T} = 2\mu\boldsymbol{S} - 2\mu(\nabla \cdot \boldsymbol{U})\boldsymbol{I}/3$, where $\boldsymbol{S} = 0.5(\nabla \boldsymbol{U} + \nabla \boldsymbol{U}^T)$ and **I** is identity matrix), **f** is gravitational force. Parameters $\mu_b$ and $\rho_b$ are bulk viscosity and density are based on the weighted average of the distribution of phase fraction:

$$\mu_b = \alpha\mu_C + (1-\alpha)\mu_D \quad (4)$$

$$\rho_b = \alpha\rho_C + (1-\alpha)\rho_D \quad (5)$$

The last term on the right-hand side of Eqn. 2, **F**$_S$, represents the continuum surface tension force (CSF) [57] and is nonzero only on the interface. This force term is defined as $\boldsymbol{F}_s = \gamma\kappa(\nabla\alpha)$ where κ is curvature ($\kappa = \nabla \cdot (\frac{\nabla\alpha}{|\nabla\alpha|})$).

We solved the momentum balance equation in conjunction with the continuity equation using the Pressure Implicit with Splitting of Operators (PISO) method [58]. In this method, the velocity field is predicted and then corrected to advance the pressure and velocity fields in time. In this work we used three PISO iterations. Once the velocity field was found, Eqn. 3 was solved to find the phase fraction. Even though Eqns. (1-3) yield velocity and phase fraction at every cell in the domain, the location of the interface needs to be identified with high resolution. To achieve this, we used a two fluid formulation where the contribution of each phase to the velocity of interface is considered, i.e.

$$\frac{\partial \alpha}{\partial \bar{t}} + \nabla \cdot (\alpha \boldsymbol{U}_C) = 0 \tag{6}$$

$$\frac{\partial (1-\alpha)}{\partial \bar{t}} + \nabla \cdot ((1-\alpha) \boldsymbol{U}_D) = 0 \tag{7}$$

where $\mathbf{U}_C$ and $\mathbf{U}_D$ are velocity vector fields of continuous and dispersed phase respectively. Here, we assumed that velocity of each phase has a contribution to the convection of interface based on their phase fraction, i.e.

$$\boldsymbol{U} = \alpha \boldsymbol{U}_C + (1-\alpha) \boldsymbol{U}_D \tag{8}$$

Equation (6) can be rearranged and used as phase fraction equation [59]:

$$\frac{\partial \alpha}{\partial \bar{t}} + \nabla \cdot (\alpha \boldsymbol{U}) - \nabla \cdot (\alpha(1-\alpha) \boldsymbol{U}_r) = 0 \tag{9}$$

where $\boldsymbol{U}_r = \boldsymbol{U}_D - \boldsymbol{U}_C$, is called as the compression velocity. Eqn. 9 has a new convective term, compared to Eqn. 3. This term is only present at the interface and vanishes in the pure phases. The compression velocity is given by:

$$\boldsymbol{U}_r = \boldsymbol{n}_f \min[C_\alpha \left|\frac{\varphi}{S_f}\right|, \max\left(\frac{|\varphi|}{|S_f|}\right)] \tag{10}$$

$\boldsymbol{n}_f$ is the cell normal flux, the ratio $\left(\frac{|\varphi|}{|S_f|}\right)$] is the magnitude of velocity where φ and $\mathbf{S}_f$ are the cell face volume flux and surface area, respectively. $C_\alpha$ is a user-specified compression factor that can vary from zero to four. Using larger compression factor results in thinner interface. However, Hoang *et al.* who studied microfluidic droplet break-up showed that compression factor greater than one causes parasitic currents at interface [44], therefore we chose $C_\alpha = 1$ in this study.

In order to assure stability and convergence of the simulation we used an adaptive time step method. At the beginning of each iteration, a new time step was calculated based on the Courant number (Co), which is defined as:

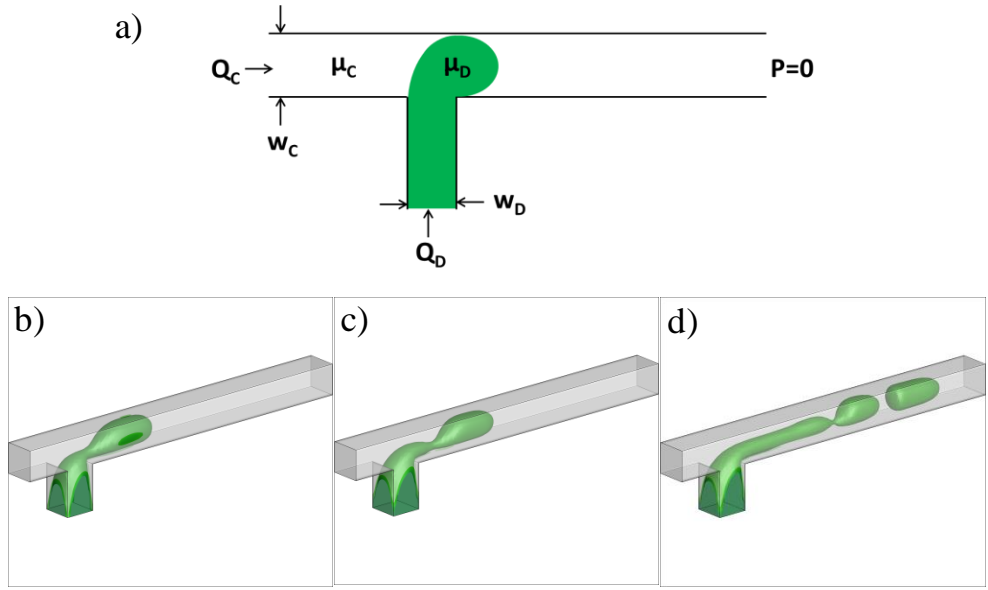

Figure 1. Drop formation at a microfluidic T-junction. (a) Schematic of the microfluidic T-junction highlighting the relevant system parameters. The definition of these parameters is provided in the main text. Representative images showing the three modes of droplet breakup (b) Squeezing, Ca=0.01 (c) Dripping, Ca=0.03 (d) Jetting, Ca=0.07. The different breakup behaviors were obtained by varying the capillary number, while keeping other system parameters fixed (Q=0.2, λ=0.1, W=1 and H =1).

$$Co = \frac{|\mathbf{U}_f \cdot \mathbf{S}_f|}{d_f \cdot \mathbf{S}_f} \Delta \bar{t} \tag{11}$$

where $d_f$ is the distance between two neighbor cells and $\mathbf{U}_f$ is the velocity at the surface of the cell and $\Delta \bar{t}$ is time step. Then based on the identified courant number, a new time step size was calculated in order to keep Co less than a predefined limit. The value of Courant number indicates how much each fluid element is displaced over one time step. For example, when Co=1 it means that, each element of the fluid moves in a distance of one grid size in one time step. Courant number was evaluated at each computational cell including bulk fluid and the interface in each iteration.

All the numerical simulations were performed in the open-source code OpenFOAM. VOF was implemented in the interFoam solver as reported by Deshpande *et al.*[60]. Several works have used interFoam for solving incompressible two phase flows [19, 44, 61-63]. Among those papers Hoang *et al.* and Nieves-Remacha *et al.* have used this solver for simulation of two phase flows at micro-scale [19, 44, 61]. In OpenFOAM, governing equations were discretized by finite volume center-based method. Boundedness of phase fraction parameter was controlled by a Total Variation Diminishing (TVD) method implemented in OpenFOAM called MULES (Multidimensional Universal Limiter with Explicit Solution) [64].

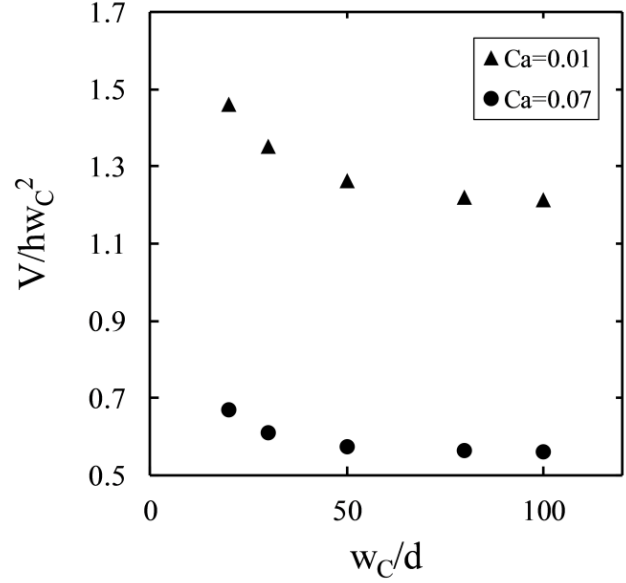

Figure 2. Effect of mesh size on the generated droplet volume in a T-junction geometry, evaluated at two different capillary numbers. The conditions of the simulation are Q=0.5, λ=1, W=0.5 and H=0.33.

## 2.2. Simulation Setup

The schematic of T-junction's geometry used in the simulation study is shown in Figure 1. The system has two inlets for continuous and dispersed phases and one outlet. We set $\alpha = 0$ at the inlet of the dispersed phase and $\alpha = 1$ at the inlet of the continuous phase. A constant velocity was set for both the inlets in the system. No slip boundary condition was applied at the walls. All the simulations were carried out for a density ratio, $\rho = 1$ and Reynolds number is Re < 0.1, therefore inertia is negligible. In this study, time is scaled by $\frac{w_D}{U_D}$.

In order to predict the values of phase fraction near the walls, i.e. the interface, the normal of interface is related to the wall's normal and tangential unit vectors:

$$n_{interface} = \cos(\theta)\, n_{wall} + \sin(\theta) t_{wall} \tag{12}$$

where $n_{interface}$ is the normal of interface, θ is the wall's static contact angle and $n_{wall}$ and $t_{wall}$ are the unit normal and unit tangential vectors to the wall, respectively. The walls of the channels were considered as fully wetted by the continuous phase; therefore we set the contact angle as zero at the walls in this study. The boundary condition at the outlet was set as atmospheric pressure.

The simulation domain was meshed using hexahedral cells. Adjacent cells to the wall were refined five times to capture lubrication film around the droplet. The domain meshing was carried out using ANSYS Gambit version 2.4.6. Since determining the size of the droplets is the main goal of this study, grid size of the mesh becomes important. We carried out numerical simulations on a representative T-junction geometry to evaluate the sensitivity of grid size (d) on the droplet volume. The width and height ratio were set to be W=0.5 and H=0.33 respectively. We performed simulations for a fixed flow rate ratio (Q=0.5) and two capillary numbers (Ca=0.01 and 0.07). We measured the droplet volume (V) by counting the number of grid points in which $\alpha \leq 0.5$. The results are shown in Fig. 2. We find that for $w_C/d > 50$, the variation in droplet size is smaller than 4%, while for $w_C/d > 75$ it's smaller than 1%. Therefore, a grid size of $w_C/d=100$ was chosen in this study.

To clarify whether the upstream and downstream channel lengths of T-junction are long enough for the flow to become fully developed, we computed the entrance length in our system. The entrance length for channel flows under laminar conditions (Re<2000) is given by[65]

$$L_e = d_h(\frac{0.6}{1+0.035Re} + 0.056Re)$$

where $d_h$ is the hydraulic diameter of the channel. We find that the maximum entrance length, corresponding to highest Re, needed to achieve a fully developed flow is just a small fraction of the upstream ($L_e$=0.11 $L_{Upstream}$) and downstream lengths ($L_e$=0.02 $L_{Downstream}$) used in the simulation. Therefore, the choice of entrance lengths used in the simulation is sufficient to obtain fully developed flow.

All the simulations were performed in parallel on a Linux cluster (Hrothgar cluster at High Performance Computing Center, Texas Tech University and Stampede cluster, Texas Advanced Computing Center) by employing 48 to 96 processors. Based on initial test simulations, we determined that in order to prevent spurious currents and nonphysical behavior of interface, courant number and interface courant number need to be kept below 0.1. Typical clock time for the formation of one droplet ranges from 22 – 40 hrs depending on the number of processors used.

## 3. Results and discussion

### 3.1. Regimes of Drop Formation

Previous studies have shown that three distinct regimes – squeezing, dripping and jetting - exist for the dispersed phase behavior in a T-junction geometry [11, 38, 66, 67]. In this section, we perform simulations to examine whether VOF can capture the different regimes of drop production. We also compare our simulation results to the experimental data of Tice *et al*.[11] who identified these regimes as a function of flow rate ratio and capillary number.

Similar to experiments, we observe the three behaviors in our VOF simulations as shown in Fig. 1. We find that in the squeezing regime, the dispersed phase blocks the main channel significantly and breakup occurs in the vicinity of the T-junction. In the dripping regime, the dispersed phase only blocks the main channel partially, and penetrates further from the T-junction and the droplets are produced at a fixed spatial location in the microchannel. In contrast to the dripping regime, in the jetting regime, we find that after the generation of the first droplet the thread of the dispersed phase continues to move forward and a second droplet is produced further downstream. At high capillary numbers, Ca > 0.1, we observed that dispersed phase co-flows with the continuous phase forming parallel stream without any droplet being formed. These different behaviors, we observed with VOF simulations are consistent with those observed in experiments [11, 27, 67], phase field [38] and lattice-Boltzmann simulations[68].

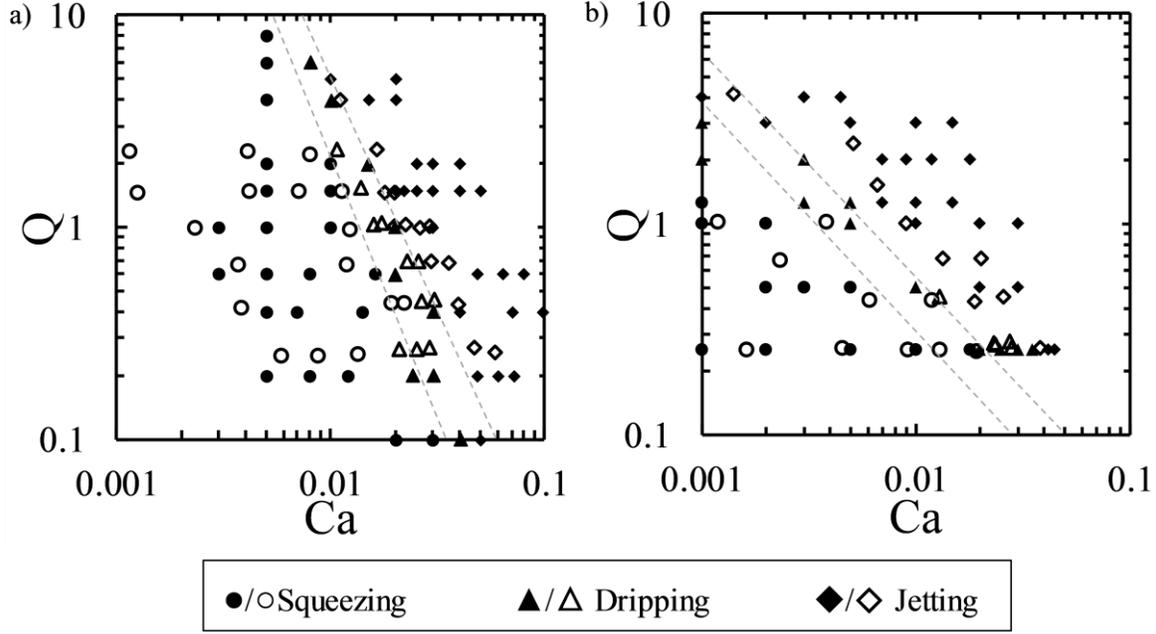

Figure 3. Regimes of drop formation in a microfluidic T-junction device with W=1, H=1. VOF simulations (closed symbols) were conducted for (a) $\lambda=0.1$ and (b) $\lambda=10$. Experimental data (open symbols) are from Tice et al.[11]. The dashed lines are drawn to guide the eye.

To test whether our VOF simulation can quantitatively predict the transitions between these regimes as a function of system parameters, we compared the results from our simulation with the experimental data of Tice et al. [11]. In their study, experiments were conducted for a geometry with W=H=1, for two different viscosity ratios $\lambda=0.1$ and 10, and for $0.1 < Q < 10$ and $10^{-3} <$ Ca $< 10^{-1}$. Simulations were performed at the same parameter values as the experiments.

Figure 3 shows the map of the regimes as a function of the two control parameters - Ca and Q. A good agreement is observed between the simulation results and the experimental data. As shown in Figure 3, for relatively small capillary number and flow rate ratio, droplets are produced in the squeezing regime. However, by increasing Ca and/or Q, the regime of drop formation changes to dripping and jetting. The dripping region is much narrower compared to squeezing and jetting. By increasing the flow rate ratio, we find that the transition from squeezing to dripping and dripping to jetting occurs at a lower capillary number. This observation indicates that at higher flow rate ratios dispersed phase penetrates much more into the main channel such that the continuous phase can break it up only in the downstream region.

By comparing results for two different viscosity ratios in Fig. 3, we observe that the boundaries defining the transition between regimes are steeper for $\lambda=10$ than $\lambda=0.1$. This means that the transition between regimes occurs at significantly smaller capillary numbers for higher viscosity ratio. When the viscous force of dispersed phase increases, it is difficult for the continuous phase to fragment it. The viscous force of dispersed phase depends on both the viscosity and velocity of dispersed phase. Therefore, for high viscosity ratios or high velocities of dispersed phase, drops are generated in jetting regime.

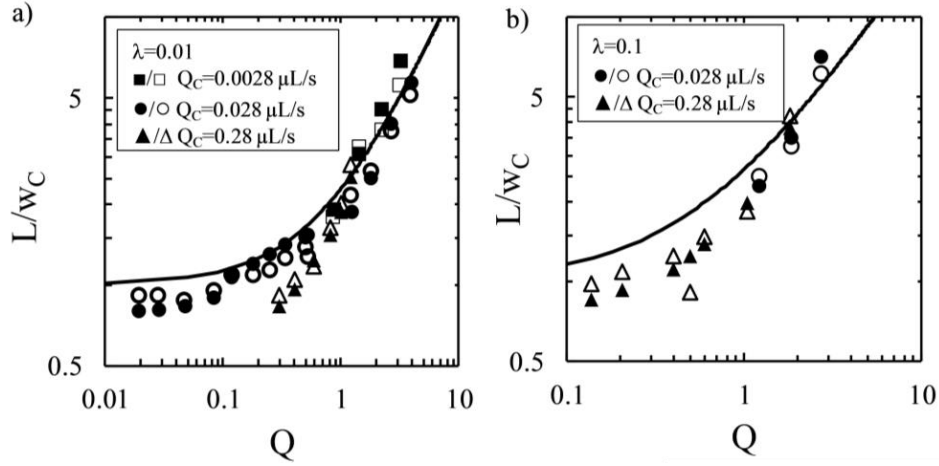

Figure 4. Effect of flow rate ratio on the size of droplets for (a) λ=0.01 and (b) 0.1. Experimental data (open symbols) are from reference [30] and the VOF data are represented by closed symbols. The geometrical parameters are W=0.5, H=0.33 and the flow rates of continuous phase are 0.0028, 0.028 and 0.28 μL/S. The prediction based on Eqn. 14 (see main text) is shown by (−).

## 3.2. Validating VOF simulation using experimental data of droplet size

In the previous section, we have quantified the regimes of drop formation in the T-junction device with VOF simulations. We note it is difficult to define sharply the transition between the regimes. Therefore, in order to benchmark the capabilities of VOF simulations more precisely we compared VOF predictions of drop size against experimental data. As mentioned earlier, we choose the three experimental data sets reported in Table 2 to validate the VOF method.

Figure 4 compares the data reported by Garstecki *et al*. and the numerical simulations in a fixed T-junction geometry at two different viscosity ratios, λ=0.01 and 0.1. The experiments were conducted at three different continuous phase flow rates, $Q_c$ = 0.0028, 0.028 and 0.28 μL/s. Garstecki *et al*. [30] plotted the droplet length normalized with the main channel width, $L/w_c$, as a function of flow rate ratio, Q. They found that the droplet length is mostly constant at low flow rate ratios but increases almost linearly at high flow rate ratios. Our simulation results also show similar trends and are in good agreement with their experimental data.

The experimental data obtained by Garstecki et al. in Fig. 4 pertains to a single T-junction geometry. To explore the capability of VOF simulations to predict droplet size generated in other T-junction geometries, we relied on the experimental study by Van Steijn *et al*. [31]. They used three different geometries with different width and height ratios, (W, H) = (0.33, 0.33), (0.67, 0.17) and

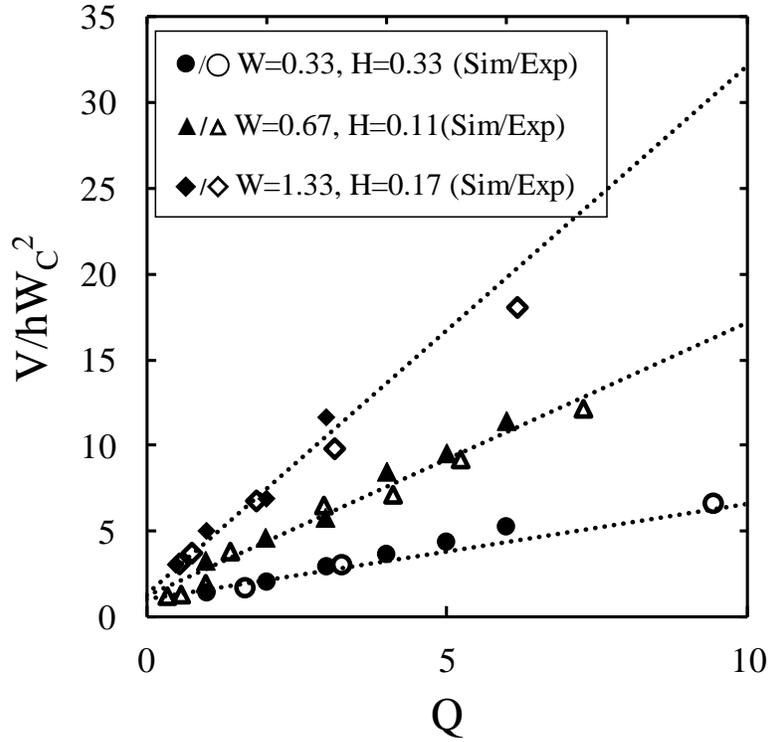

Figure 5. Effect of T-junction geometry on the generated droplet size for $\lambda=0.1$. Experimental data (open symbols) are from reference [31]. VOF data (closed symbols) were obtained for $Ca \sim O(10^{-3})$. The dashed lines are drawn using equations in Figure 2 of reference [31].

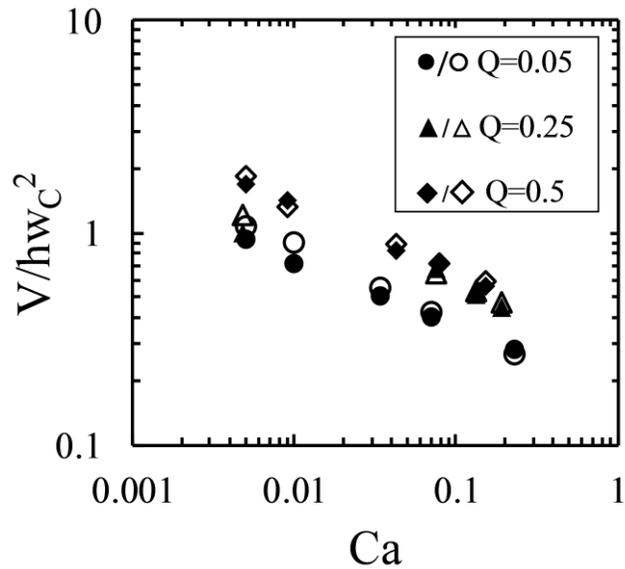

Figure 6. Effect of capillary number on the size of droplets for $W=1$, $H=0.5$ and $\lambda=0.01$. Experimental data (open symbols) are from reference [29] and VOF data is shown by closed symbols.

(1.33, 0.11). We used the same geometries in our study and the same viscosity ratio of $\lambda=0.1$. We varied the flow rate ratio ($0.2 < Q < 6$) by fixing the continuous phase flow rate and changing the flow rate of dispersed phase. Van Steijn *et al.* plotted normalized droplet volume, $V/hw_c^2$, as a function of flow rate ratio for the three mentioned geometries (see Fig. 5). They found that the width of the side channel influences the volume of droplet significantly. At a fixed flow rate ratio, the volume of produced droplet in a T-junction with a wider side channel is larger. For example, at Q = 5 the droplet volume in T-junction with W=1.33 is four times larger than in W=0.33. As shown in Fig. 5, our simulation results predict the same trend of data as the experimental findings.

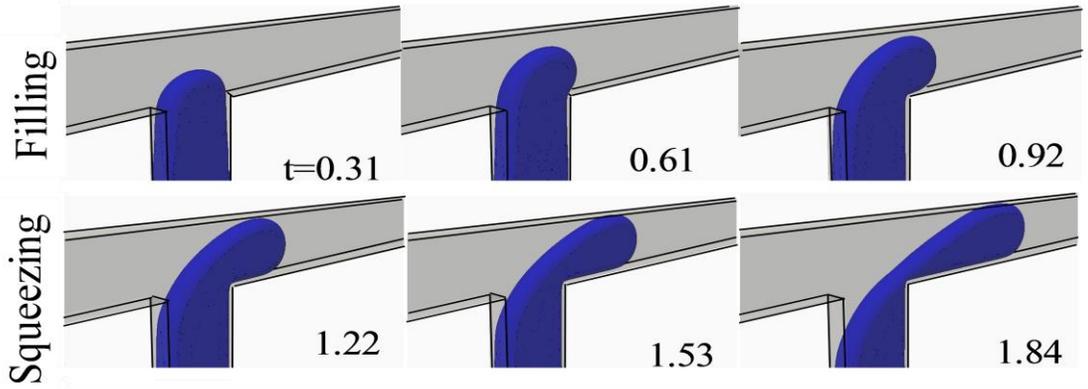

Figure 7. Images from the VOF simulation (for Q=0.8, Ca=0.05, λ=0.1, W=1 and H=0.33) showing the droplet formation process, highlighting the filling and squeezing stages. During the filling stage, the dispersed phase occupies the main channel. Because of the blockage of the main channel by the dispersed phase, the upstream pressure builds, squeezing the neck, which connects the dispersed phase to droplet.

The experimental data of Garstecki *et al.* and van Stein *et al.* pertain mostly to low capillary numbers (Ca ~ $O(10^{-3})$). We therefore conducted VOF simulations at high capillary numbers and compared the results with that of the experimental data reported by Christopher *et al.*[29]. They employed one set of fluids, λ=0.01 and one geometry, W=1 and H=0.33 in their experimental study. They fixed the flow rate ratios at, Q=0.05, 0.25 and 0.5, and changed the capillary number by varying inlet velocity of continuous phase. We performed numerical simulations at the same parameter values as experiments. As shown in Figure 6, our numerical findings are consistent with the experimental data. By increasing the capillary number for any fixed value of flow rate ratio the droplet volume decreases. On the other hand, by increasing flow rate ratio at fixed capillary number the volume of droplet increases.

Overall, comparison of simulation results with the three sets of experimental data reveals that our VOF methodology and choice of simulation parameters (e.g. grid size, compression factor, Courant number) are suitable for making measurements of droplet size in microfluidic T-junction devices.

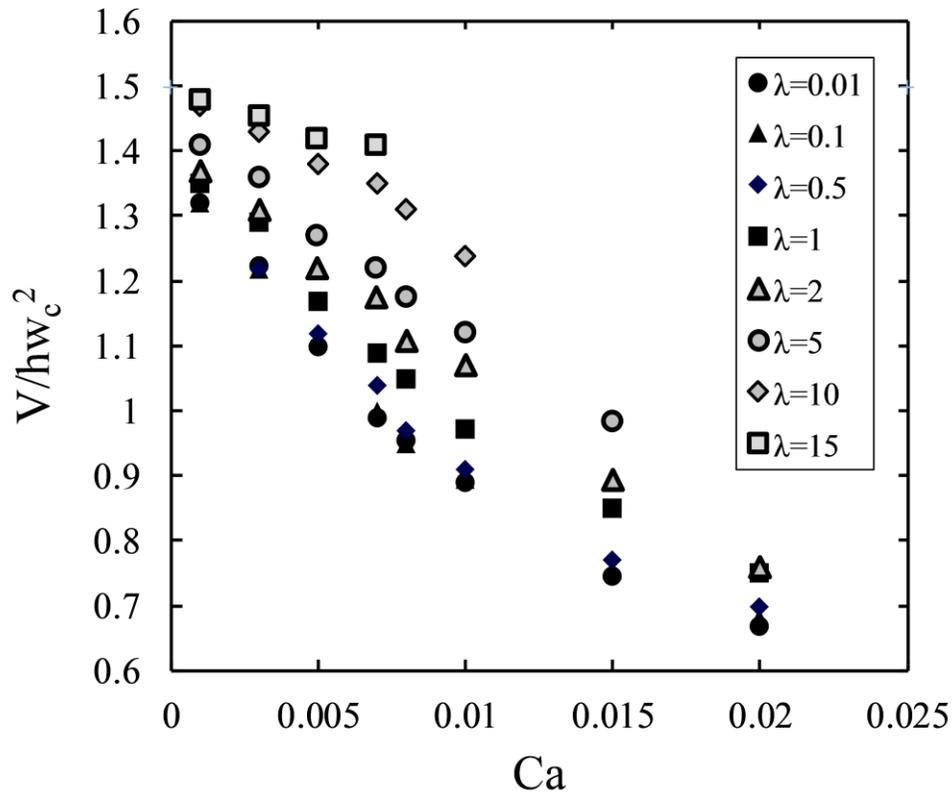

Figure 8. Dependence of droplet volume on the capillary number for different viscosity ratios. Other parameters were fixed at Q = 0.3, W = 1 and H = 0.33.

### 3.3. Importance of fluid viscosity ratio on droplet size

In this section, we discuss a popular model for predicting droplet size at low capillary number in T-junction devices, as originally proposed by Garstecki *et al.*[30]. Given that this simple model ignores the effect of fluid viscosity ratio, we subsequently study the influence of viscosity ratio using VOF simulations and highlight the need to incorporate this parameter into analytical predictions of droplet size.

At low capillary numbers, Garstecki *et al.* assume that a drop is produced in a two-stage process. In the first stage, the dispersed phase starts to enter the main channel and occupies a portion of the channel. This stage is called the filling stage. In the second stage, called the squeezing stage, the neck which connects the drop to the dispersed phase is squeezed by the continuous phase, ultimately pinching-off the dispersed phase. Figure 7 shows snapshots from the VOF simulation depicting these two stages of droplet generation.

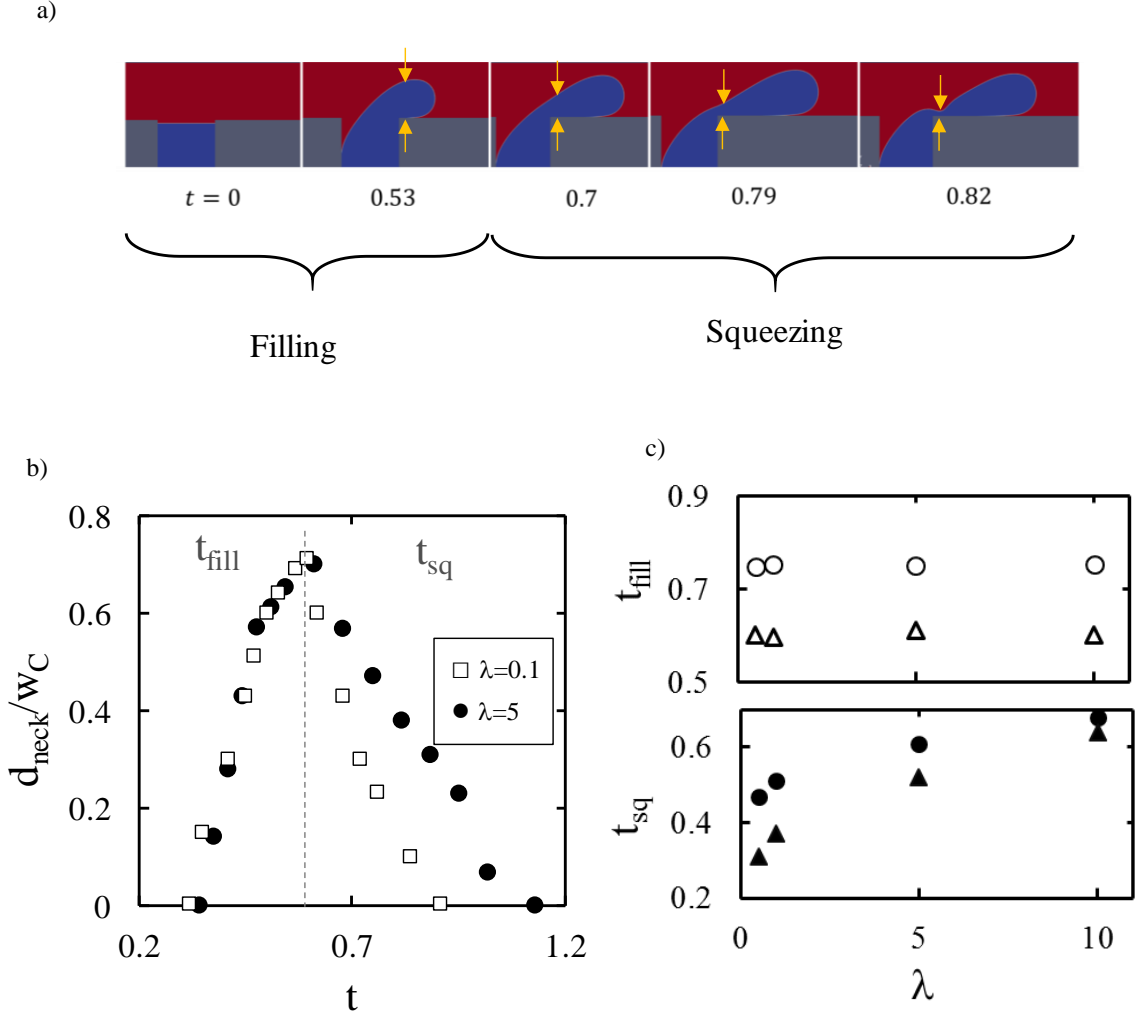

Figure 9. The effect of viscosity ratio on the thickness of dispersed phase and the timescales during droplet formation. The simulations were performed for Ca = 0.01, Q=0.3, W=1 and H=0.33 (a) Snapshots of mid-plane view of the dispersed phase as a function of time. Yellow arrows represent thickness of the neck, $d_{neck}$, over time. Regions corresponding to filling and squeezing are highlighted (b) Time-evolution of dispersed phase thickness for two viscosity ratios of λ=0.1 and 5. The dashed vertical line demarcates the duration of filling and squeezing stages. (c) Plot of squeezing and filling times as a function of viscosity ratios for Ca = 0.003 (circles) and Ca=0.01 (triangles).

Using this conceptual picture, Garstecki *et al.* proposed scaling arguments to obtain a prediction for droplet size [30]. They argued that during the filling stage, the extent to which the dispersed phase fills the main channel is approximately equal to the width of the main channel, $w_c$. In the squeezing step, the droplet is being filled by an amount equal to $U_D t_{sq}$ where $U_D$ is the dispersed phase velocity and $t_{sq}$ is the time needed to squeeze and rupture the dispersed phase finger. By adding these two contributions to the overall droplet length, we have $L = w_C + U_D t_{sq}$. Assuming that the dispersed phase is being squeezed by a rate proportional to the continuous phase inlet velocity ($U_C$) gives an estimate of $t_{sq} = \frac{d_C}{U_C}$, where $d_C$ is an undetermined length scale. Using this estimate of $t_{sq}$, the equation for droplet length can be rewritten as

$$\frac{L}{w_C} = 1 + \beta Q \tag{14}$$

where $\beta = d_C/w_C \sim O(1)$ is a fitting parameter that depends on the T-junction geometry. For example, for the data in Fig. 4, $\beta = 1.13 \pm 0.16$ and $1.67 \pm 0.41$ for $\lambda = 0.01$ and $0.1$ respectively. The effect of the T-junction geometry was explicitly considered by Van Steijn *et al.*, as reflected by the data of Fig. 5.

The low capillary number model of Garstecki *et al.* relates the droplet size to simply the flow rate ratio and does not consider fluid viscosity ratio. Likewise, the compilation of studies in Tables 1 and 2 did not cover a broad range of viscosity ratios and were conducted with fluids for which $\lambda \leq 1$. We therefore investigated the importance of viscosity ratio by varying $\lambda$ from 0.01 to 15. In the simulations, $\lambda$ was varied by changing the viscosity of dispersed phase and keeping the continuous phase viscosity constant. Additionally, simulations were conducted at a fixed flow rate ratio (Q=0.3) in a device geometry with W=1 and H=0.33. We chose this flow rate ratio as it allows a wider access into the squeezing and dripping regimes. However, at higher viscosity ratios ($\lambda$=10, 15), we obtained limited data, because the operating window for squeezing/dripping regime is small as discussed earlier in section 3.1. The capillary number corresponding to these simulations was $0.001 < Ca < 0.02$.

Fig. 8 shows the droplet size as a function of Ca for different viscosity ratios. For all viscosity ratios, by increasing the Ca, the size of droplet decreases suggesting that increasing the viscous stress of continuous phase produces smaller droplets. However, the reduction in droplet size with Ca is stronger for $\lambda < 1$ than for $\lambda > 1$. In addition, we find that at a given Ca, the size of droplets does not vary appreciably when $\lambda < 1$, while it increases when $\lambda > 1$.

To understand the observations on the effect of viscosity ratio, we followed the time evolution of the width of the dispersed phase neck ($d_{neck}$) during the breakup process, as shown in Fig. 9a. Here $d_{neck}$ is measured at the channel location where ultimately the pinch-off occurs. Fig.9b shows how the non-dimensional neck thickness, $d_{neck}/w_D$, varies with time for $\lambda = 0.1$ and 5. As expected, we observe that the neck thickness increases and then decreases with a maximum. We define the duration in which the neck thickness continues to increase and then decrease as $t_{fill}$ and $t_{sq}$ respectively, since these two time scales effectively correspond to the filling and squeezing stages in the droplet generation process.

From the data in Fig. 9b, we observe that at a fixed capillary number, the duration of filling stage for the two viscosity ratios is the same while the duration of the squeezing stage is different. In fact, as shown in Fig. 9c, when explored across the entire viscosity range ($0.01 \leq \lambda \leq 15$) for varying capillary number, we find $t_{fill}$ is independent of viscosity ratio and is larger when the capillary number is low. In contrast, $t_{sq}$ increases with viscosity ratio, but decreases with increase in capillary number. The observed dependence of $t_{fill}$ and $t_{sq}$ on Ca and $\lambda$ help to explain our simulations results shown in Fig. 8. Since both $t_{fill}$ and $t_{sq}$ decrease with Ca, the drop volume decreases with increase in Ca. At a fixed Ca, since $t_{sq}$ increases with viscosity ratio, while $t_{fill}$ is constant we observe larger droplet size for higher viscosity ratio.

### 3.4. Improved model for predicting droplet size

Our results thus far highlight that the viscosity ratio affects the dispersed phase breakup behavior and that the squeezing duration increases with viscosity ratio, leading to larger droplet sizes. The drop breakup model of Garstecki *et al.* discussed earlier (see Sec. 3.3) estimates the squeezing time as $t_{sq} = \frac{d_C}{U_C}$. Clearly, the effect of viscosity ratio is missing in this estimate of squeezing time. Here we address this gap and develop an improved model for predicting the generated droplet size.

In our model, similar to previous work [30, 31], we assume the overall drop volume is resulting from volumetric contributions during the filling and squeezing stages, *i.e.*

$$V = V_{fill} + V_{sq} \tag{15}$$

To estimate the volume of the droplet at the end of the filling period, we use Eqn. 16, which was derived by Christopher *et al.*[29]. Eqn.16 was obtained by conducting a force balance that includes continuous phase shear stress, upstream fluid pressure and Laplace pressure jumps across the front and rear of the droplet.

$$\left(1 - \sqrt{\frac{V_{fill}}{hw_C^2}}\right)^3 = Ca\sqrt{\frac{V_{fill}}{hw_C^2}} \tag{16}$$

Eqn.16 is consistent with our simulation result that $t_{fill}$ is independent of viscosity ratio but dependent on capillary number (c.f. Fig. 9). In addition, as $Ca \to 0$, Eqn.16 predicts that the non-dimensional fill volume $\frac{V_{fill}}{hw_C^2} \to 1$, which is consistent with the low capillary number model of Garstecki *et al.*

To estimate $V_{sq}$, we determine the overall squeezing time ($T_{sq}$) by summing the break-up time of a viscous thread ($t_b$) and the squeezing time considered by Garstecki *et al.*, i.e.

$$T_{sq} = t_{sq} + t_b \tag{17}$$

To determine $t_b$, we consider the break-up dynamics of Newtonian filaments that has been investigated in several studies [69-73]. These studies report the viscous filament break-up time to be $t_b = k\frac{\mu_D d_C}{\gamma}$, with $6 < k < 33$. This time scale results from the competition between surface tension trying to shrink the liquid filament and the viscous stresses in the filament opposing it.

Incorporating the expressions for $t_{sq}$ and $t_b$ into Eqn. 17 we obtain

$$T_{sq} = \frac{d_C}{U_C}[1 + k\lambda Ca] \tag{18}$$

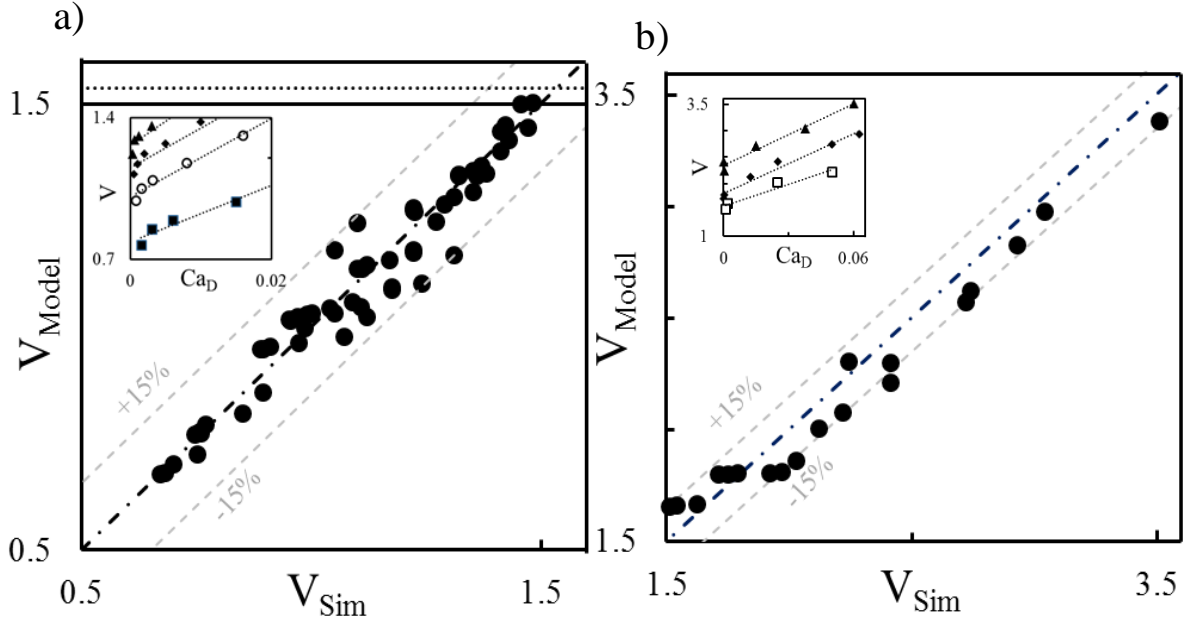

Figure 10. Parity plot showing the prediction of the improved model incorporating viscosity ratio, versus the simulation data for a) W=1, H=0.33 and 0.01≤λ≤15 and b) W=1, H=1 and λ=0.1 and 10. In this plot, V is the volume of the droplet normalized by $hw_C^2$. Models predictions reported by Garstecki *et al.* and Van Steijn *et al.* are shown by horizontal solid and dotted lines, respectively [30,31]. The grey lines indicate 15% deviation from the expected true value. The insets show the model predictions (dotted lines) as a function of $Ca_D$ for different Ca values: ▲ 0.003, ◆ 0.005, ○ 0.008, □ 0.01 and ■ 0.015.

Given the volumetric contribution to the drop size during the squeezing period is $V_{sq} = T_{sq}Q_D$, we have

$$\frac{V_{sq}}{hw_C^2} = \beta Q + \xi Ca_D \tag{19}$$

where $\xi = k \frac{w_D d_C}{w_C^2}$.

Solving Eqns. (15), (16) and (19) together gives the final prediction for the generated droplet volume in T-junction devices taking into account the viscosity ratio of the fluids.

In Figure 10, we compare the predictions of our improved model with the simulation data across a wide range of viscosity ratio and for two geometries with different aspect ratios H=0.33 and 1. For the first geometry, H=0.33, the viscosity ratio varies between 0.01 and 15 and for the second geometry, H=1, we have λ=0.1 and 10. Two sets of fit parameters ξ and β were used for these two geometries. We find excellent agreement between our model predictions and the simulation data capturing different viscosity ratios. For each set of data with the same Ca, we found the best-fit values of ξ. For the first geometry, H=0.33, this value varies between 7 to 36 for the entire set of data with an average of 21.96±8.6. For the second geometry, H=1, the best-fit value of ξ varies between 8 and 19 with an average of 14.8±3.69. These finding are consistent with the range of $k$ values reported in the literature ($5 < k < 33$) since we have $\frac{w_D d_C}{w_C^2} \sim O(1)$. According to McKinley and Tripathi, the k-value depends on the longitudinal stress in the viscous thread, which is a function of the axial curvature of the filament[69]. In our study, the rate at which the dispersed phase is injected into the droplet changes when the capillary number is varied. This can cause the axial curvature of the dispersed phase finger to vary with capillary number and therefore it can result in variability of k-values, as observed when we generated the parity plot (Fig. 10). We also found the best fit parameter $\beta = 1.18 \pm 0.4$ for the first geometry, and $\beta = 1.4 \pm 0.77$ for the second geometry. These values are reasonable estimates since we get $\beta = 1.8$ and 1.15 for the first and second geometry respectively, when evaluated with the theoretical model proposed by Van Steijn *et al.* that includes the effect of T-junction geometry [31].

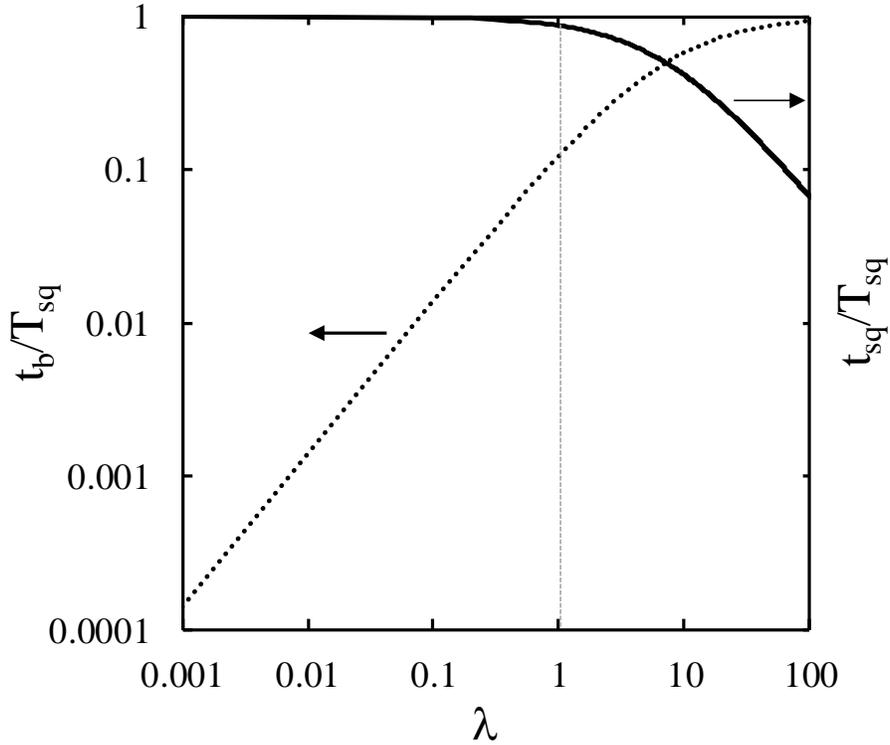

Figure 11. Dependence of normalized contributions of breakup and squeezing time-scales with viscosity ratio. The dashed vertical line indicates that the contribution of the breakup time to the overall squeezing time is small for $\lambda \leq 1$.

An important outcome of our analysis is that it explains why at a given (low) capillary number, the generated drop volumes are the same for $\lambda < 1$ but varies for $\lambda > 1$ (see Fig. 8). As shown in Fig. 11, when we plot the contributions of $t_{sq}$ and $t_b$ to the overall squeezing time as a function of viscosity ratio, we find that for $\lambda < 1$, the contribution of viscous thread breakup time is negligible compared to the squeezing time. This implies that the produced droplets will have the same size for $\lambda < 1$. Alternatively, we observe that for $\lambda > 1$, the breakup time contributes more making the drop volumes dependent on viscosity ratio.

## Conclusions

We implemented the volume-of-fluid (VOF) method for investigating droplet generation in microfluidic T-junction devices. We examined the different regimes of dispersed phase behavior and found the VOF simulations to be in good agreement with experimental maps of the regimes. We also performed simulations at low and high capillary numbers, as well as in different T-junction geometries. The predicted droplet size was in good agreement with experimental reports in the literature. Taken together, these results validate our VOF methodology and choice of simulation parameters.

Our simulation results show that for $\lambda > 1$, the droplet size depends on viscosity ratio. Previous theoretical models have not considered this dependence of viscosity ratio on the generated droplet size. We developed an improved model and incorporated this dependence by including the break-up time needed to fragment the high-viscosity dispersed phase. This improved model predicts successfully the influence of drop volume on capillary number as well as viscosity ratio.

More broadly, we find that VOF is a useful tool for droplet-based microfluidics and has the potential to reveal new insights into the fluid physics of drop formation and behavior, which can have important technological applications.

## Acknowledgments

We thank the National Science Foundation (CAREER Grant No. 1150836 and AIR-TT Grant No.1445070) for supporting this work. We are grateful to Jerzy Blawzdziewicz and Michael Loewenberg for useful discussions.

## References

1. H. Song, D. L. Chen and R. F. Ismagilov,"Reactions in droplets in microfluidic channels", Angew Chem Int Ed Engl **45** (44), 7336-7356 (2006).
2. S. Y. Teh, R. Lin, L. H. Hung and A. P. Lee,"Droplet microfluidics", Lab Chip **8** (2), 198-220 (2008).
3. X. C. I. Solvas and A. deMello,"Droplet microfluidics: recent developments and future applications", Chem Commun **47** (7), 1936-1942 (2011).
4. H. N. Joensson and H. A. Svahn,"Droplet Microfluidics-A Tool for Single-Cell Analysis", Angew Chem Int Edit **51** (49), 12176-12192 (2012).
5. R. Seemann, M. Brinkmann, T. Pfohl and S. Herminghaus,"Droplet based microfluidics", Rep Prog Phys **75** (1) (2012).
6. A. Dewan, J. Kim, R. H. McLean, S. A. Vanapalli and M. N. Karim,"Growth kinetics of microalgae in microfluidic static droplet arrays", Biotechnology and bioengineering **109** (12), 2987-2996 (2012).
7. J. Avesar, T. B. Arye and S. Levenberg,"Frontier microfluidic techniques for short and long-term single cell analysis", Lab on a Chip **14** (13), 2161-2167 (2014).
8. M. A. Khorshidi, P. K. P. Rajeswari, C. Wählby, H. N. Joensson and H. A. Svahn,"Automated analysis of dynamic behavior of single cells in picoliter droplets", Lab on a Chip **14** (5), 931-937 (2014).
9. H. Song, J. D. Tice and R. F. Ismagilov,"A microfluidic system for controlling reaction networks in time", Angewandte Chemie **115** (7), 792-796 (2003).

10. A. M. Huebner, C. Abell, W. T. Huck, C. N. Baroud and F. Hollfelder,"Monitoring a reaction at submillisecond resolution in picoliter volumes", Analytical chemistry **83** (4), 1462-1468 (2011).
11. J. D. Tice, A. D. Lyon and R. F. Ismagilov,"Effects of viscosity on droplet formation and mixing in microfluidic channels", Analytica chimica acta **507** (1), 73-77 (2004).
12. P. Paik, V. K. Pamula and R. B. Fair,"Rapid droplet mixers for digital microfluidic systems", Lab on a Chip **3** (4), 253-259 (2003).
13. W. S. Wang and S. A. Vanapalli,"Mechanisms of mass transport during coalescence-induced microfluidic drop dilution", Physical Review Fluids **1** (6), 064001 (2016).
14. M. J. Fuerstman, A. Lai, M. E. Thurlow, S. S. Shevkoplyas, H. A. Stone and G. M. Whitesides,"The pressure drop along rectangular microchannels containing bubbles", Lab on a Chip **7** (11), 1479-1489 (2007).
15. S. S. Bithi and S. A. Vanapalli,"Behavior of a train of droplets in a fluidic network with hydrodynamic traps", Biomicrofluidics **4** (4), 044110 (2010).
16. C. N. Baroud, F. Gallaire and R. Dangla,"Dynamics of microfluidic droplets", Lab on a Chip **10** (16), 2032-2045 (2010).
17. S. A. Vanapalli, A. G. Banpurkar, D. van den Ende, M. H. Duits and F. Mugele,"Hydrodynamic resistance of single confined moving drops in rectangular microchannels", Lab on a Chip **9** (7), 982-990 (2009).
18. A. Leshansky, S. Afkhami, M.-C. Jullien and P. Tabeling,"Obstructed breakup of slender drops in a microfluidic T junction", Physical review letters **108** (26), 264502 (2012).
19. D. A. Hoang, L. M. Portela, C. R. Kleijn, M. T. Kreutzer and V. van Steijn,"Dynamics of droplet breakup in a T-junction", Journal of Fluid Mechanics **717** (2013).
20. D. Link, S. L. Anna, D. Weitz and H. Stone,"Geometrically mediated breakup of drops in microfluidic devices", Physical review letters **92** (5), 054503 (2004).
21. M. Sun and S. A. Vanapalli,"Generation of chemical concentration gradients in mobile droplet arrays via fragmentation of long immiscible diluting plugs", Analytical chemistry **85** (4), 2044-2048 (2013).
22. S. S. Bithi, W. S. Wang, M. Sun, J. Blawzdziewicz and S. A. Vanapalli,"Coalescing drops in microfluidic parking networks: A multifunctional platform for drop-based microfluidics", Biomicrofluidics **8** (3), 034118 (2014).
23. B. Bhattacharjee and S. A. Vanapalli,"Electrocoalescence based serial dilution of microfluidic droplets", Biomicrofluidics **8** (4), 044111 (2014).
24. P. Janssen, M. Baron, P. Anderson, J. Blawzdziewicz, M. Loewenberg and E. Wajnryb,"Collective dynamics of confined rigid spheres and deformable drops", Soft Matter **8** (28), 7495-7506 (2012).
25. S. S. Bithi and S. A. Vanapalli,"Collective dynamics of non-coalescing and coalescing droplets in microfluidic parking networks", Soft matter **11** (25), 5122-5132 (2015).
26. T. Ward, M. Faivre, M. Abkarian and H. A. Stone,"Microfluidic flow focusing: Drop size and scaling in pressure versus flow-rate-driven pumping", Electrophoresis **26** (19), 3716-3724 (2005).
27. J. Nunes, S. Tsai, J. Wan and H. Stone,"Dripping and jetting in microfluidic multiphase flows applied to particle and fibre synthesis", Journal of physics D: Applied physics **46** (11), 114002 (2013).
28. G. Christopher and S. Anna,"Microfluidic methods for generating continuous droplet streams", Journal of Physics D: Applied Physics **40** (19), R319 (2007).
29. G. F. Christopher, N. N. Noharuddin, J. A. Taylor and S. L. Anna,"Experimental observations of the squeezing-to-dripping transition in T-shaped microfluidic junctions", Phys Rev E Stat Nonlin Soft Matter Phys **78** (3 Pt 2), 036317 (2008).
30. P. Garstecki, M. J. Fuerstman, H. A. Stone and G. M. Whitesides,"Formation of droplets and bubbles in a microfluidic T-junction-scaling and mechanism of break-up", Lab Chip **6** (3), 437-446 (2006).
31. V. van Steijn, C. R. Kleijn and M. T. Kreutzer,"Predictive model for the size of bubbles and droplets created in microfluidic T-junctions", Lab Chip **10** (19), 2513-2518 (2010).

32. V. van Steijn, M. T. Kreutzer and C. R. Kleijn,"μ-PIV study of the formation of segmented flow in microfluidic T-junctions", Chemical Engineering Science **62** (24), 7505-7514 (2007).
33. J. Xu, S. Li, J. Tan and G. Luo,"Correlations of droplet formation in T-junction microfluidic devices: from squeezing to dripping", Microfluidics and Nanofluidics **5** (6), 711-717 (2008).
34. T. Glawdel, C. Elbuken and C. L. Ren,"Droplet formation in microfluidic T-junction generators operating in the transitional regime. I. Experimental observations", Physical Review E **85** (1) (2012).
35. T. Glawdel, C. Elbuken and C. L. Ren,"Droplet formation in microfluidic T-junction generators operating in the transitional regime. II. Modeling", Physical Review E **85** (1), 016323 (2012).
36. T. Fu, Y. Ma, D. Funfschilling, C. Zhu and H. Z. Li,"Squeezing-to-dripping transition for bubble formation in a microfluidic T-junction", Chemical engineering science **65** (12), 3739-3748 (2010).
37. J. Sivasamy, T.-N. Wong, N.-T. Nguyen and L. T.-H. Kao,"An investigation on the mechanism of droplet formation in a microfluidic T-junction", Microfluidics and nanofluidics **11** (1), 1-10 (2011).
38. M. De Menech, P. Garstecki, F. Jousse and H. Stone,"Transition from squeezing to dripping in a microfluidic T-shaped junction", journal of fluid mechanics **595**, 141-161 (2008).
39. H. Yang, Q. Zhou and L.-S. Fan,"Three-dimensional numerical study on droplet formation and cell encapsulation process in a micro T-junction", Chemical Engineering Science **87**, 100-110 (2013).
40. Y. Yan, D. Guo and S. Wen,"Numerical simulation of junction point pressure during droplet formation in a microfluidic T-junction", Chemical Engineering Science **84**, 591-601 (2012).
41. S. Van der Graaf, T. Nisisako, C. Schroen, R. Van Der Sman and R. Boom,"Lattice Boltzmann simulations of droplet formation in a T-shaped microchannel", Langmuir **22** (9), 4144-4152 (2006).
42. M. N. Kashid, A. Renken and L. Kiwi-Minsker,"CFD modelling of liquid–liquid multiphase microstructured reactor: Slug flow generation", Chemical Engineering Research and Design **88** (3), 362-368 (2010).
43. H. H. Liu and Y. H. Zhang,"Droplet formation in a T-shaped microfluidic junction", J Appl Phys **106** (3) (2009).
44. D. A. Hoang, V. van Steijn, L. M. Portela, M. T. Kreutzer and C. R. Kleijn,"Benchmark numerical simulations of segmented two-phase flows in microchannels using the Volume of Fluid method", Computers & Fluids **86**, 28-36 (2013).
45. L. Sang, Y. Hong and F. Wang,"Investigation of viscosity effect on droplet formation in T-shaped microchannels by numerical and analytical methods", Microfluidics and nanofluidics **6** (5), 621-635 (2009).
46. C. W. Hirt and B. D. Nichols,"Volume of fluid (VOF) method for the dynamics of free boundaries", Journal of computational physics **39** (1), 201-225 (1981).
47. D. Gueyffier, J. Li, A. Nadim, R. Scardovelli and S. Zaleski,"Volume-of-fluid interface tracking with smoothed surface stress methods for three-dimensional flows", Journal of Computational Physics **152** (2), 423-456 (1999).
48. D. J. Benson,"Volume of fluid interface reconstruction methods for multi-material problems", Applied Mechanics Reviews **55** (2), 151-165 (2002).
49. B. Van Wachem and A.-E. Almstedt,"Methods for multiphase computational fluid dynamics", Chemical Engineering Journal **96** (1), 81-98 (2003).
50. V. Badalassi, H. Ceniceros and S. Banerjee,"Computation of multiphase systems with phase field models", Journal of Computational Physics **190** (2), 371-397 (2003).
51. R. Folch, J. Casademunt, A. Hernández-Machado and L. Ramirez-Piscina,"Phase-field model for Hele-Shaw flows with arbitrary viscosity contrast. I. Theoretical approach", Physical Review E **60** (2), 1724 (1999).
52. M. De Menech,"Modeling of droplet breakup in a microfluidic T-shaped junction with a phase-field model", Physical Review E **73** (3), 031505 (2006).
53. S. Chen and G. D. Doolen,"Lattice Boltzmann method for fluid flows", Annual review of fluid mechanics **30** (1), 329-364 (1998).
54. D. Yu, R. Mei, L.-S. Luo and W. Shyy,"Viscous flow computations with the method of lattice Boltzmann equation", Progress in Aerospace Sciences **39** (5), 329-367 (2003).

55. C. K. Aidun and J. R. Clausen,"Lattice-Boltzmann method for complex flows", Annual review of fluid mechanics **42**, 439-472 (2010).
56. R. R. Nourgaliev, T.-N. Dinh, T. Theofanous and D. Joseph,"The lattice Boltzmann equation method: theoretical interpretation, numerics and implications", International Journal of Multiphase Flow **29** (1), 117-169 (2003).
57. J. Brackbill, D. B. Kothe and C. Zemach,"A continuum method for modeling surface tension", Journal of computational physics **100** (2), 335-354 (1992).
58. H. Rusche, Imperial College London (University of London), 2003.
59. E. Berberović, N. P. van Hinsberg, S. Jakirlić, I. V. Roisman and C. Tropea,"Drop impact onto a liquid layer of finite thickness: Dynamics of the cavity evolution", Physical Review E **79** (3), 036306 (2009).
60. S. S. Deshpande, L. Anumolu and M. F. Trujillo,"Evaluating the performance of the two-phase flow solver interFoam", Computational science & discovery **5** (1), 014016 (2012).
61. M. J. Nieves-Remacha, L. Yang and K. F. Jensen,"OpenFOAM Computational Fluid Dynamic Simulations of Two-Phase Flow and Mass Transfer in an Advanced-Flow Reactor", Ind Eng Chem Res **54** (26), 6649-6659 (2015).
62. A. Q. Raeini, M. J. Blunt and B. Bijeljic,"Modelling two-phase flow in porous media at the pore scale using the volume-of-fluid method", Journal of Computational Physics **231** (17), 5653-5668 (2012).
63. F. Raees, D. Van der Heul and C. Vuik, Report No. 1389-6520, 2011.
64. O. OpenCFD,"The Open Source CFD Toolbox", User Guide, OpenCFD Ltd (2009).
65. C. J. Pipe, T. S. Majmudar and G. H. McKinley,"High shear rate viscometry", Rheologica Acta **47** (5-6), 621-642 (2008).
66. A. Gupta and R. Kumar,"Effect of geometry on droplet formation in the squeezing regime in a microfluidic T-junction", Microfluidics and Nanofluidics **8** (6), 799-812 (2010).
67. T. Thorsen, R. W. Roberts, F. H. Arnold and S. R. Quake,"Dynamic pattern formation in a vesicle-generating microfluidic device", Physical Review Letters **86** (18), 4163-4166 (2001).
68. A. Gupta, S. S. Murshed and R. Kumar,"Droplet formation and stability of flows in a microfluidic T-junction", Appl Phys Lett **94** (16), 164107 (2009).
69. G. H. McKinley and A. Tripathi,"How to extract the Newtonian viscosity from capillary breakup measurements in a filament rheometer", Journal of Rheology (1978-present) **44** (3), 653-670 (2000).
70. D. T. Papageorgiou,"On the breakup of viscous liquid threads", Physics of Fluids (1994-present) **7** (7), 1529-1544 (1995).
71. J. Eggers,"Universal pinching of 3D axisymmetric free-surface flow", Physical Review Letters **71** (21), 3458 (1993).
72. M. P. Brenner, J. R. Lister and H. A. Stone,"Pinching threads, singularities and the number 0.0304", Physics of Fluids (1994-present) **8** (11), 2827-2836 (1996).
73. L. E. Rodd, T. P. Scott, J. J. Cooper-White and G. H. McKinley,"Capillary Break-up Rheometry of Low-Viscosity Elastic Fluids", Applied Rheology **15** (1) (2005).